\documentclass[sigconf]{acmart}
\newcommand{\ownesty}{\textit{Ownesty}}

\copyrightyear{2020} 
\acmYear{2020} 
\setcopyright{acmcopyright}\acmConference[ICPC '20]{28th International Conference on Program Comprehension}{October 5--6, 2020}{Seoul, Republic of Korea}
\acmBooktitle{28th International Conference on Program Comprehension (ICPC '20), October 5--6, 2020, Seoul, Republic of Korea}

\setcopyright{none}

\begin{document}

\title{Ownership at Large\vspace{50\in}}

\subtitle{Open Problems and Challenges in Ownership Management\vspace{50\in}}

\author{John Ahlgren, Maria Eugenia Berezin, Kinga Bojarczuk, Elena Dulskyte, Inna Dvortsova,
Johann George, Natalija Gucevska, Mark Harman, Shan He, Ralf L\"ammel, Erik Meijer, Silvia Sapora, and Justin Spahr-Summers}
\authornote{Author order is alphabetical.
Contact author: Ralf L\"ammel (\texttt{rlaemmel@acm.org}).
This paper appears in Proceedings of 28th International Conference on Program Comprehension, ICPC 2020.
The subject of the paper is covered by the contact author's keynote at the same conference.}
\affiliation{Facebook Inc.\vspace{150\in}\vspace{150\in}}

\begin{abstract}
Software-intensive organizations rely on large numbers of software assets of different types, e.g., source-code files, tables in the data warehouse, and software configurations. Who is the most suitable owner of a given asset changes over time, e.g., due to reorganization and individual function changes. New forms of automation can help suggest more suitable owners for any given asset at a given point in time. By such efforts on ownership health, accountability of ownership is increased. The problem of finding the most suitable owners for an asset is essentially a program comprehension problem: how do we automatically determine who would be best placed to understand, maintain, evolve (and thereby assume {\em ownership} of) a given asset. This paper introduces the Facebook \ownesty{} system, which uses a combination of ultra large scale data mining and machine learning and has been deployed at Facebook as part of the company's ownership management approach. \ownesty{} processes many millions of software assets (e.g., source-code files) and it takes into account workflow and organizational aspects. The paper sets out open problems and challenges on ownership for the research community with advances expected from the fields of software engineering, programming languages, and machine learning.
\vspace{150\in}
\end{abstract}


\maketitle

\renewcommand{\shortauthors}{John Ahlgren et al.}


\section{Introduction}

Managing software asset ownership in any organization is important. Many pressing industrial concerns such as security, reliability, and integrity depend crucially on well-defined ownership so that there are clear lines of responsibility for maintenance tasks, code review, incident response, and others. Ownership management requires and connects research on a wide variety of topics including program comprehension, and more generally, software engineering, programming languages, and machine learning.

In this paper, when we refer to (software) `asset' we include entities as diverse as source-code files, tables in the data warehouse, and software configurations. When we refer to the `owner' of an asset, we mean this term in a broad sense: a set of people who take responsibility for the asset. The set can be singleton, but may also be a group or sub-organization. The owner can also vary depending on purpose~--~such as code review versus incident response. If the set was ever empty, the asset is {\em unowned}. Standard processes, e.g., based on escalation, are typically in place to rule out unowned assets, as they would clearly be a cause for concern. A more nuanced question is the one of `ownership health', i.e., whether each asset is attributed to the `most suitable' owner.  Who is the most suitable owner of a given asset changes over time, e.g., due to reorganization and individual function changes. Ownership health give rises to interesting research problems and challenges. 

Attributing assets to owners and measuring ownership health encompasses a combination of static and dynamic properties of the software assets themselves, the workflows for developing and managing the assets, and the structures of the organization that possesses the assets. As such, the problem of ownership draws on topics from a diverse set of research fields and previously studied problem domains, such as \emph{global software engineering}~\cite{EbertKP16a,EbertKP16b,DafoulasMAAC17} at the highest level of abstraction through to \emph{program dependence analysis} \cite{dbmh:advances,silva:slice-vocab,weiser:slicing79} at the lowest level of abstraction. 

The paper outlines the authors' work at Facebook on the problem of ownership management with a focus on ultra large
scale data mining and machine learning, subject to collaboration with other teams focusing on additional aspects such as tooling and workflow integration. This work has resulted in the \ownesty{} system, which is introduced in Section~\ref{S:system}.

There remain many open challenges and problems that need to be addressed in the more specific context of, for example, reverse engineering, architecture recovery, mining software repositories, process mining, and interpretable models. None of these challenges and problems are specific to the Facebook setting and, in fact, much of the progress in this area can be expected to be achieved in the context of research on open-source ecosystems. Therefore, the paper describes, in Section~\ref{S:open}, a set of challenges and open problems for an ongoing research agenda on modern ownership management.


\section{The \ownesty{} System}
\label{S:system}

Let us briefly describe the \ownesty{} system for ownership management, as developed and used at Facebook.


\subsection{Vocabulary of Ownership Management}

The term \emph{asset} refers to any sort of entity that is a part of a system or is possessed by a company of interest. (We skip over hardware here for simplicity.) Examples of assets are these: a file in the repository for a system, a database that is part of the system, a VM to run the system, or a configuration of the VM. Ownership is also lifted to compound or distributed entities such as components, products, apps, or the scattered implementation of a logging feature.

The term \emph{owner candidate} refers to any sort of individual or group entity which is associated with the system (or company) of interest and could possibly be accountable for any number of assets in this scope. In the work on \ownesty{} at Facebook, we deal with a few types of owner candidates: \emph{individual owner}, \emph{team} (supported by a director), \emph{reporting team} (supported by a manager), and \emph{oncall rotation} (some sort of response team type). There are a few more obscure types that we skip here. In the engineering practice, the types \emph{individual owner} and \emph{oncall rotation} are particularly important.

We assume a special part of a system: its \emph{asset-to-owner attribution mapping} or just \emph{attribution}, which maps assets to owner candidates, which are thus referred to as \emph{owners}. Individuals or processes with appropriate permissions may modify the mapping. In particular, when an asset is mapped to a new owner, then this may be referred to as \emph{ownership transfer}. The main purpose of a system like \ownesty{} is to recommend suitable owners and thereby also to validate ownership health, i.e., the suitability of the currently attributed owners. To this end, machine learning and heuristics are leveraged. Humans may be in the loop for the purpose of confirmation, also depending on the degree of confidence for the available recommendations.


\begin{figure}[t!]
\includegraphics[width=.87\columnwidth]{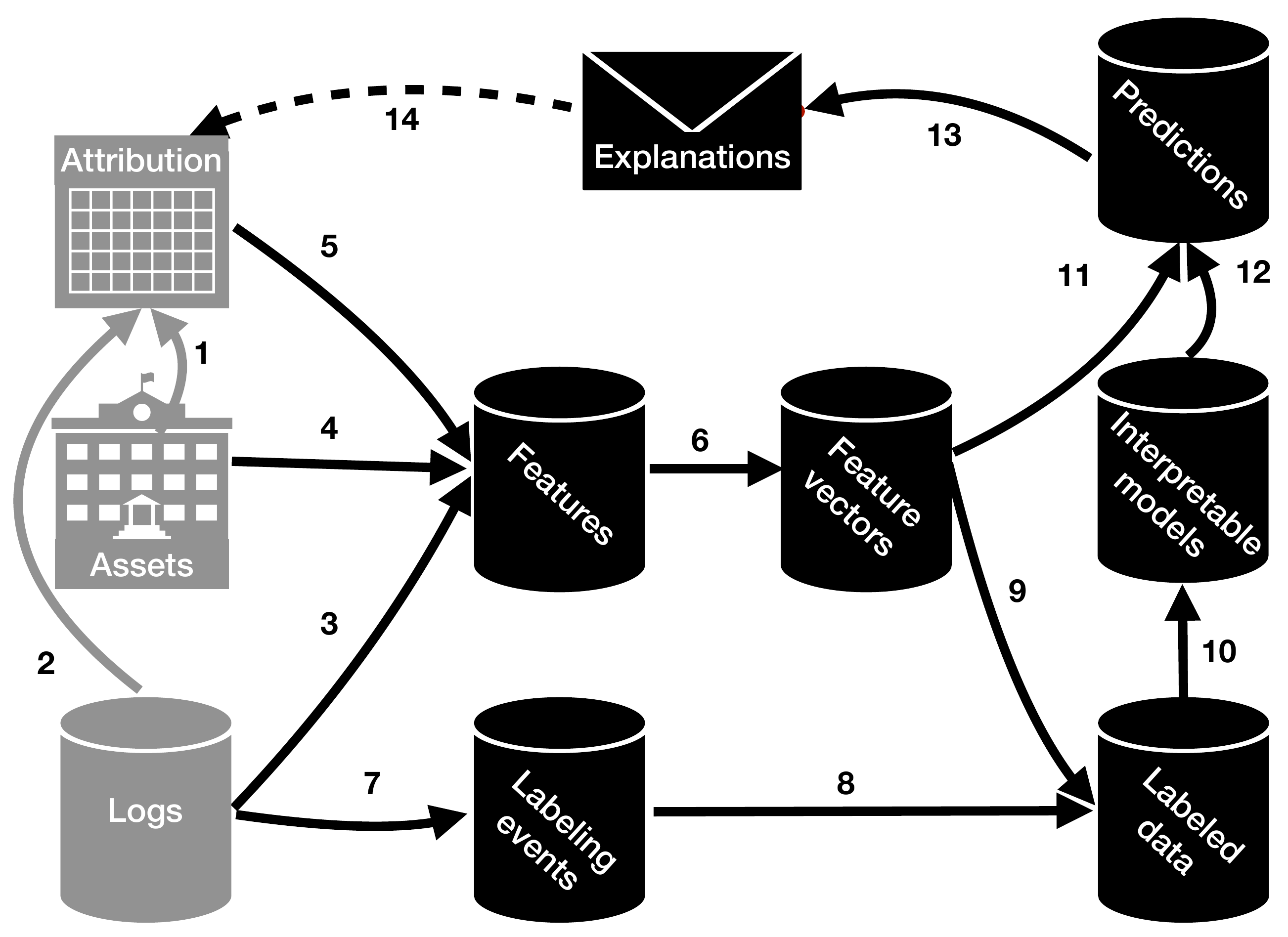}
\caption{Overall \ownesty{} data flow for ownership recommendation. {\small(See the text for the numbered arrows. The data flow starts from the asset-to-owner attribution mapping on the left.)}}
\vspace{-22\in}
\label{F:data-flows}
\Description{See caption.}
\vspace{-32\in}
\end{figure}


\subsection{The ML Architecture}

In \autoref{F:data-flows}, the arrows denote data flow (computation). The gray shapes and arrows (see on the left) exist regardless of \ownesty{}. Several of the arrows are supported by metadata, which we do not further detail here for brevity.

The gray arrows on the left express that the asset-to-owner attribution mapping is partially encoded in the assets themselves such as by annotation within files or a metastore for tables, in which case extraction can be applied to assets (1) or possibly to logs (2) that record the owners `in action'.

\ownesty{} extracts features from the available logs (3) that record some relevant form of interactions between assets and owner candidates. (For instance, a log for a database admin tool would record who was taking what administrative action when.) This is a data and feature engineering challenge because of the plethora of logs and the fact that they were not designed with ownership management in mind. Feature extraction also involves assets and attribution (4--5), e.g., features obtained by source-code analysis. (For instance, we may extract a feature regarding an oncall annotation from a build file.) The individually extracted features are composed into feature vectors (6)~--~these are specific to the asset type.

\ownesty{} leverages supervised learning and thus relies on labeled data for positive and negative attribution. To this end, so-called `labeling events' are extracted from the logs (7), e.g., events that recorded reliable human decisions to accept or reject owner recommendations for attributing assets to owner candidates. The labeled data for training and test is then obtained by joining labeled events with the feature vectors for those events (8--9).

We build interpretable models and provide prediction sets (10--12) for the various asset types. Interpretable or explainable models (e.g., basic decision trees or linear models lifted to scoring systems) are essential because the predictions and the underlying models need to be understood by humans.

Subject to further metadata (e.g., documentation for the features), predictions are mapped to actionable `explanations' and surfaced through project/ownership management tooling (13) so that humans in the loop can accept or reject, thereby modifying the asset-to-owner attribution mapping (14) (and providing more labeled data eventually).


\subsection{Ownership at Large}

In this section we describe the scale of ownership management at Facebook, giving some key metrics we use to characterize the asset-owner space covered.

\paragraph*{Number of asset types} We distinguish a few hundreds of types. Of course, not every type calls for advanced heuristics or ML for ownership management. The number of asset types is an artifact of the kind of distinctions made. For instance, we do not simply use the type `database table', but we distinguish different storage engines, as they need to be addressed in different ways, e.g., in terms of ownership signal, in fact, features.

\paragraph*{Number of assets of type $t$} This depends on $t$, of course. For instance, 
there are many millions of files under version control; there are millions of tables based on different storage engines; there are several 100k assets for scheduled pipelines in the data warehouse. Even a type with just 10k+ assets may benefit from advanced heuristics or ML, if precision is high and the investment is outweighed by the resulting savings from the automation.

\paragraph*{Number of owner types} As discussed, there are two particularly important ones: individual owners and oncall owners. 

\paragraph*{Number of owner candidates of type $t$} The number of individual owner candidates translates to the number of employees or engineers or yet other appropriately defined subsets of employees at Facebook; we typically look at several 1k or 10k individual owner candidates. The number of oncall rotations is less than 10k.

\paragraph*{Number of (shortlisted) owner candidates for a given asset $a$} It is often possible, subject to heuristics based on key features, to narrow down to a short list of (3-100) owner candidates that are at all plausible for the asset $a$ and that need to be ranked thus. 

\paragraph*{Daily churn for asset type $t$} (That is, the number of assets of type $t$ that are added, deleted, or changed per day.) Such churn is relevant, as it may affect ownership, but see the next metric for deeper insight. For instance, the daily churn for source-code files in one of the bigger mono repos of Facebook is around 100k.

\paragraph*{Daily owner churn for asset type $t$} (That is, the number of assets of type $t$ where the owner is changed per day.) Such churn is relevant as heuristics / ML are supposed to automate (recommend, validate) these changes. For instance, for an important asset type for scheduled pipelines in the data warehouse with about 100k assets, the aggregated daily owner churn over the last 4 months is about 10k while there were several days with several hundreds of affected assets~--~this is a consequence of prioritized efforts on ownership management at Facebook, based on \ownesty{} or otherwise.



\section{Open Problems and Challenges}
\label{S:open}

We lay out a number of research areas around ownership by describing the open problems and challenges in these areas and referring to related work to capture the state of the art.


\subsection{Heterogeneity of Owned Assets}

Ownership recommendation~--~especially when focusing on code assets~--~is related to \emph{code authorship attribution}~\cite{KalgutkarKGSM19}, as relevant, for example, for detecting malicious or plagiarized code, subject to stylometry methods and the overarching assumption of distinctive writing style to be viewed as a fingerprint.

Ownership recommendation bears also similarities with \emph{reviewer recommendation}~\cite{Lipcak018} which aims at recommending reviewers for new patches (diffs, commits) based on a model built from past patches and possibly other data. For instance, one may extract features such as filenames, module, author, lines deleted, added, number of braces and train a Bayesian Network for recommendations~\cite{JeongLZY09}. Reviewer recommendation focuses on code assets~--~here: files and patches under the regime of version control and code review. We mention in passing that there are many challenges of automating reviewer recommendations at scale~\cite{AsthanaKBBBMMA19}, e.g., the need for load balancing so that reviewers are not overloaded.

Ownership recommendation needs to address the heterogeneity of asset types such as database tables and software configurations in addition to just code. Even just the `code type' breaks down into many different subtypes based on language and purpose. Each asset type necessitates specific features. Accordingly, a generic core is needed to be reused across different types and `patterns' are needed to help onboarding new types. All these features are to be organized and standardized in a manner to convey and leverage similarities across asset types. Further, the multitude of models and the underlying computations need to be managed in an efficient and robust manner.

Within each asset there resides a wealth of information that can be used to determine a suitable owner. Such information has been the topic of study in the program comprehension community for many years.  For example, program slicing~\cite{korel:understanding}, concept assignment~\cite{mhetal:cosac}, and search-based optimization~\cite{mh:icpc-keynote}, as well as many other analysis techniques, have all been used to investigate structural components of a software asset to support programmers' comprehension of the asset.  The same kind of information can be used to provide features to machine learning, the aim of which is to identify owner candidates who are best-placed to understand the asset in question.


\subsection{Dependency Awareness}

We cannot look at assets in isolation, but we need to leverage and respect various kinds of dependencies. Let us draw again inspiration from reviewer recommendation with an instance of heterogeneous dependencies, in this case, between regular and library code~\cite{RahmanRC16} where such dependencies are aggregated over all developer contributions, thereby essentially aggregating developer experiences which can be considered in addition to `blame'-based information for finding reviewers. (The cited work relies on a form of token extraction applied to regular and library code; it uses then cosine similarity for comparing aggregated experience of developers with the `required'  experience for patches in need of a review.)

More generally, we need to take into account build dependencies (e.g., a file being generated from another file), usage dependencies (e.g., a database table being consumed by a pipeline), feature mapping (e.g., a logging configuration being associated with a product feature), product mapping (e.g., a collection of assets being shared across products, subject to per-product owners), and requirements to assets mapping. Several of the mentioned dependencies are also version-/variant-specific.

Research is hence needed to integrate ownership management, with various software engineering aspects such as 
feature location~\cite{DitRGP13},
software variability~\cite{BergerRNABCW13},
package management and reuse~\cite{BenelallamHSBB19},
build management~\cite{KonatEV18},
traceability recovery~\cite{TraceabilityBook,RempelMKP13},
change impact analysis~\cite{ren:chianti,PoshyvanykMFG09,LiSLZ13}, and
software ecosystems~\cite{LunguRL10,JansenC12,Manikas16,ManikasH13,LiuHWL17}.

Existing dependency analyses need to be further generalized to apply better to heterogeneous assets and the problem of attributing assets to owners. For instance, provenance or lineage may be aimed at dependencies for data assets while information flow may be aimed at program assets, but combinations or generalizations of such methods are needed to cover the asset types that occur together in practice~\cite{AcarACP12,CheneyAA11,SunGDY16}.


\subsection{Workflow and Organizational Aspects}

Attribution of assets to owners also needs to take into account interactions of owner candidates with the assets. In Section~\ref{S:system}, we already discussed the essential use of system logs for ownership recommendation such as interpreting the logged used of a database admin tool as an ownership signal. Let us receive inspiration again from the area of reviewer recommendation, where additional developer-workflow data such as reviewer activity for commits or commenting in an issue tracker are leveraged to identify and rank reviewer candidates~\cite{YuWYW16}.

Clearly, ownership recommendation requires a generalization of the analysis of interactions to account for the different types of assets and owner candidates and diverse forms of interaction. Ultimately, the identification of the most suitable owners for assets relies on a deeper understanding of the involved workflows of engineers. For instance, we may take into account project management-based workflow constraints~\cite{CataldoMRH09}. In this manner, we enter the realm of \emph{process mining} or workflow modeling and encounter the challenging notion of case ID recovery~\cite{NezhadBSCA08,GoelBW12,workitem}.

Human-to-asset and human-to-human interaction and collaboration does not only exercise workflow aspects; it also relates to organizational aspects of ownership management. In this manner, we enter the realm of \emph{global software engineering}~\cite{EbertKP16a,EbertKP16b,DafoulasMAAC17,WangR16}. The systematic extraction, integration, and interpretation of all the diverse ownership-related signal (per-asset data, asset dependencies, workflows, organizational structures) calls for \emph{knowledge management}~\cite{Girard15}, subject to a dedicated knowledge graph~\cite{hogan2020knowledge}.

To be more concrete, organizational structure may support better understanding of ownership in that, for example, the health of a particular attribution of an asset to an owner may depend on past, recent, and upcoming changes to teams (`reorganizations') or individual roles. This may suggest future research to revise existing (software engineering) concepts. For example, consider change-impact analysis~\cite{ren:chianti,PoshyvanykMFG09,LiSLZ13}, which needs to be advanced to take into account organizational aspects~--~the impact of a change depends not {\em only} on the forward slice of the change locations, but also the owners of those affected assets in the forward slice.

Human aspects of ownership and their interplay with technical aspects provide a rich area of future research. We can expect Computer Supported Collaborative Working (CSCW)~\cite{andriessen:arguing} and Crowdsourced Software Engineering (CSE)~\cite{ke:jss-survey} to have a role to play here. Tools for CSCW can be developed or adapted to support ownership, while CSE can contribute a ground-truth approach for ownership decisions used in machine learning.


\subsection{Understandable Recommendations}

It is important for recommended owners to be `understandable', thereby entering the realm of interpretable or explainable models in machine learning~\cite{Rudin19}, giving rise to the following options.

Ideally, the ownership model is directly interpretable, as in the case of `plain' decision trees with some limit on the depth (such as 5). We can also use scoring systems based on supersparse linear integer models~\cite{RudinU18}, even though they require extra effort to deal with correlated features. (We currently favor decision tree-based algorithms in \ownesty, but also consider embeddings.) One may also commence with an indirectly interpretable model using, for example, permutation-based feature importance~\cite{AltmannTSL10}.

If we were to give up on interpretable models, we can still maintain that individual predictions (owners) can be directly explained. This is possible, for example, when decision trees (e.g., random forest or gradient boosting) are used. In addition, prediction-specific feature importance~\cite{LundbergL17} can be taken into account. (Adding some sophistication, one can also explain predictions by an interpretation around a given prediction~\cite{Ribeiro0G16}.)

When black-box models are used (e.g., embeddings with deep learning), 
individual predictions can be still explained by using counterfactuals by means of perturbing input features. For instance, an explanation can take the form ``Had you touched the file in the last 2 days, you would have been recommended as owner''.

The following domain-specific constraints challenge the provision of interpretable and explainable models for ownership recommendation; dedicated research is needed thus:

\begin{itemize}

\item The attribution relationship between assets and owner candidates may be intrinsically inconclusive. That is, some assets may be hard to associate with very suitable owners because, for example, the most suitable candidates may just have left the team or the company. Also, some assets may associate with several similarly suitable candidates, which is also problematic in terms of acting on such recommendations; see the next item.

\item The process of communicating, discussing, and deciding on ownership is a social one. For instance, ownership recommendations may be subject to rejection, delegation, and escalation~--~these decisions are not solely based on facts and the resulting limits of feature engineering and explainable predictions need to be explored. (Compare this with image recognition: Human subjects will typically agree on how to distinguish cats and dogs.)
    
\item The introduction of rigorous ownership management is a process as opposed to the installment of a recommendation system. The side effect of a project like \ownesty{} is that one provides highly structured and aggregated information that would not be available otherwise. Those who need to accept or reject recommendations may start to take a dependency on data they would not have had available before. This may lead to `concept drift' that needs to be addressed by the ML approach.

\end{itemize}


\section{Conclusion}
\label{S:concl}

This paper characterizes the general notion of ownership management
and the specific aspects of using ownership recommendation for
attributing assets to owners and measuring the health of any such
attribution for large and complex projects and systems. The
recommendation of suitable owners and the assessment of ownership
health relies on data extracted from assets (per-asset data as well as asset
dependencies), workflows and organizational structures. We
hope to stimulate interest and activity in this exciting area.  We
have introduced the Facebook \ownesty{} system to illustrate the
practical industrial relevance of the accompanying ownership research
agenda.  We also set out open problems and challenges and their
relationships to existing research activities and communities.  We are
keen to collaborate with the research communities working on software
engineering, programming languages, and machine learning on these open
problems and challenges.

\balance

\pagebreak

\bibliography{paper}


\begin{thebibliography}{46}


\ifx \showCODEN    \undefined \def \showCODEN     #1{\unskip}     \fi
\ifx \showDOI      \undefined \def \showDOI       #1{#1}\fi
\ifx \showISBNx    \undefined \def \showISBNx     #1{\unskip}     \fi
\ifx \showISBNxiii \undefined \def \showISBNxiii  #1{\unskip}     \fi
\ifx \showISSN     \undefined \def \showISSN      #1{\unskip}     \fi
\ifx \showLCCN     \undefined \def \showLCCN      #1{\unskip}     \fi
\ifx \shownote     \undefined \def \shownote      #1{#1}          \fi
\ifx \showarticletitle \undefined \def \showarticletitle #1{#1}   \fi
\ifx \showURL      \undefined \def \showURL       {\relax}        \fi
\providecommand\bibfield[2]{#2}
\providecommand\bibinfo[2]{#2}
\providecommand\natexlab[1]{#1}
\providecommand\showeprint[2][]{arXiv:#2}

\bibitem[\protect\citeauthoryear{Acar, Ahmed, Cheney, and Perera}{Acar
  et~al\mbox{.}}{2012}]%
        {AcarACP12}
\bibfield{author}{\bibinfo{person}{Umut~A. Acar}, \bibinfo{person}{Amal Ahmed},
  \bibinfo{person}{James Cheney}, {and} \bibinfo{person}{Roly Perera}.}
  \bibinfo{year}{2012}\natexlab{}.
\newblock \showarticletitle{A Core Calculus for Provenance}. In
  \bibinfo{booktitle}{\emph{Principles of Security and Trust - First
  International Conference, {POST} 2012, Held as Part of the European Joint
  Conferences on Theory and Practice of Software, {ETAPS} 2012}}
  \emph{(\bibinfo{series}{LNCS})}, Vol.~\bibinfo{volume}{7215}.
  \bibinfo{publisher}{Springer}, \bibinfo{pages}{410--429}.
\newblock


\bibitem[\protect\citeauthoryear{Altmann, Tolosi, Sander, and Lengauer}{Altmann
  et~al\mbox{.}}{2010}]%
        {AltmannTSL10}
\bibfield{author}{\bibinfo{person}{Andr{\'{e}} Altmann}, \bibinfo{person}{Laura
  Tolosi}, \bibinfo{person}{Oliver Sander}, {and} \bibinfo{person}{Thomas
  Lengauer}.} \bibinfo{year}{2010}\natexlab{}.
\newblock \showarticletitle{Permutation importance: a corrected feature
  importance measure}.
\newblock \bibinfo{journal}{\emph{Bioinformatics}} \bibinfo{volume}{26},
  \bibinfo{number}{10} (\bibinfo{year}{2010}), \bibinfo{pages}{1340--1347}.
\newblock


\bibitem[\protect\citeauthoryear{Andriessen, Baker, and Suthers}{Andriessen
  et~al\mbox{.}}{2013}]%
        {andriessen:arguing}
\bibfield{author}{\bibinfo{person}{Jerry Andriessen}, \bibinfo{person}{Michael
  Baker}, {and} \bibinfo{person}{Dan~D Suthers}.}
  \bibinfo{year}{2013}\natexlab{}.
\newblock \bibinfo{booktitle}{\emph{Arguing to learn: Confronting cognitions in
  computer-supported collaborative learning environments}}.
  Vol.~\bibinfo{volume}{1}.
\newblock \bibinfo{publisher}{Springer Science \& Business Media}.
\newblock


\bibitem[\protect\citeauthoryear{Asthana, Kumar, Bhagwan, Bird, Bansal,
  Maddila, Mehta, and Ashok}{Asthana et~al\mbox{.}}{2019}]%
        {AsthanaKBBBMMA19}
\bibfield{author}{\bibinfo{person}{Sumit Asthana}, \bibinfo{person}{Rahul
  Kumar}, \bibinfo{person}{Ranjita Bhagwan}, \bibinfo{person}{Christian Bird},
  \bibinfo{person}{Chetan Bansal}, \bibinfo{person}{Chandra~Shekhar Maddila},
  \bibinfo{person}{Sonu Mehta}, {and} \bibinfo{person}{B. Ashok}.}
  \bibinfo{year}{2019}\natexlab{}.
\newblock \showarticletitle{WhoDo: automating reviewer suggestions at scale}.
  In \bibinfo{booktitle}{\emph{Proceedings of the {ACM} Joint Meeting on
  European Software Engineering Conference and Symposium on the Foundations of
  Software Engineering, {ESEC/SIGSOFT} {FSE} 2019}}.
  \bibinfo{publisher}{{ACM}}, \bibinfo{pages}{937--945}.
\newblock


\bibitem[\protect\citeauthoryear{Benelallam, Harrand, Soto{-}Valero, Baudry,
  and Barais}{Benelallam et~al\mbox{.}}{2019}]%
        {BenelallamHSBB19}
\bibfield{author}{\bibinfo{person}{Amine Benelallam}, \bibinfo{person}{Nicolas
  Harrand}, \bibinfo{person}{C{\'{e}}sar Soto{-}Valero},
  \bibinfo{person}{Benoit Baudry}, {and} \bibinfo{person}{Olivier Barais}.}
  \bibinfo{year}{2019}\natexlab{}.
\newblock \showarticletitle{The maven dependency graph: a temporal graph-based
  representation of maven central}. In \bibinfo{booktitle}{\emph{Proceedings of
  the 16th International Conference on Mining Software Repositories, {MSR}
  2019}}. \bibinfo{publisher}{{IEEE} / {ACM}}, \bibinfo{pages}{344--348}.
\newblock


\bibitem[\protect\citeauthoryear{Berger, Rublack, Nair, Atlee, Becker,
  Czarnecki, and Wasowski}{Berger et~al\mbox{.}}{2013}]%
        {BergerRNABCW13}
\bibfield{author}{\bibinfo{person}{Thorsten Berger}, \bibinfo{person}{Ralf
  Rublack}, \bibinfo{person}{Divya Nair}, \bibinfo{person}{Joanne~M. Atlee},
  \bibinfo{person}{Martin Becker}, \bibinfo{person}{Krzysztof Czarnecki}, {and}
  \bibinfo{person}{Andrzej Wasowski}.} \bibinfo{year}{2013}\natexlab{}.
\newblock \showarticletitle{A survey of variability modeling in industrial
  practice}. In \bibinfo{booktitle}{\emph{The Seventh International Workshop on
  Variability Modelling of Software-intensive Systems, VaMoS '13}}.
  \bibinfo{publisher}{{ACM}}, \bibinfo{pages}{7:1--7:8}.
\newblock


\bibitem[\protect\citeauthoryear{Binkley and Harman}{Binkley and
  Harman}{2004}]%
        {dbmh:advances}
\bibfield{author}{\bibinfo{person}{David Binkley} {and} \bibinfo{person}{Mark
  Harman}.} \bibinfo{year}{2004}\natexlab{}.
\newblock \showarticletitle{A Survey of Empirical Results on Program Slicing}.
\newblock \bibinfo{journal}{\emph{Advances in Computers}}  \bibinfo{volume}{62}
  (\bibinfo{year}{2004}), \bibinfo{pages}{105--178}.
\newblock


\bibitem[\protect\citeauthoryear{Cataldo, Mockus, Roberts, and
  Herbsleb}{Cataldo et~al\mbox{.}}{2009}]%
        {CataldoMRH09}
\bibfield{author}{\bibinfo{person}{Marcelo Cataldo}, \bibinfo{person}{Audris
  Mockus}, \bibinfo{person}{Jeffrey~A. Roberts}, {and}
  \bibinfo{person}{James~D. Herbsleb}.} \bibinfo{year}{2009}\natexlab{}.
\newblock \showarticletitle{Software Dependencies, Work Dependencies, and Their
  Impact on Failures}.
\newblock \bibinfo{journal}{\emph{{IEEE} Trans. Software Eng.}}
  \bibinfo{volume}{35}, \bibinfo{number}{6} (\bibinfo{year}{2009}),
  \bibinfo{pages}{864--878}.
\newblock


\bibitem[\protect\citeauthoryear{Cheney, Ahmed, and Acar}{Cheney
  et~al\mbox{.}}{2011}]%
        {CheneyAA11}
\bibfield{author}{\bibinfo{person}{James Cheney}, \bibinfo{person}{Amal Ahmed},
  {and} \bibinfo{person}{Umut~A. Acar}.} \bibinfo{year}{2011}\natexlab{}.
\newblock \showarticletitle{Provenance as dependency analysis}.
\newblock \bibinfo{journal}{\emph{Mathematical Structures in Computer Science}}
  \bibinfo{volume}{21}, \bibinfo{number}{6} (\bibinfo{year}{2011}),
  \bibinfo{pages}{1301--1337}.
\newblock


\bibitem[\protect\citeauthoryear{Cleland{-}Huang, Gotel, and
  Zisman}{Cleland{-}Huang et~al\mbox{.}}{2012}]%
        {TraceabilityBook}
\bibfield{editor}{\bibinfo{person}{Jane Cleland{-}Huang}, \bibinfo{person}{Olly
  Gotel}, {and} \bibinfo{person}{Andrea Zisman}} (Eds.).
  \bibinfo{year}{2012}\natexlab{}.
\newblock \bibinfo{booktitle}{\emph{Software and Systems Traceability}}.
\newblock \bibinfo{publisher}{Springer}.
\newblock


\bibitem[\protect\citeauthoryear{Dafoulas, Maia, Ali, Augusto, and
  Cabrera}{Dafoulas et~al\mbox{.}}{2017}]%
        {DafoulasMAAC17}
\bibfield{author}{\bibinfo{person}{Georgios~A. Dafoulas},
  \bibinfo{person}{Cristiano Maia}, \bibinfo{person}{Almaas Ali},
  \bibinfo{person}{Juan~Carlos Augusto}, {and} \bibinfo{person}{Victor~Lopez
  Cabrera}.} \bibinfo{year}{2017}\natexlab{}.
\newblock \showarticletitle{Understanding Collaboration in Global Software
  Engineering {(GSE)} Teams with the Use of Sensors: Introducing a Multi-sensor
  Setting for Observing Social and Human Aspects in Project Management}. In
  \bibinfo{booktitle}{\emph{2017 International Conference on Intelligent
  Environments, {IE} 2017}}. \bibinfo{publisher}{{IEEE}},
  \bibinfo{pages}{114--121}.
\newblock


\bibitem[\protect\citeauthoryear{Dit, Revelle, Gethers, and Poshyvanyk}{Dit
  et~al\mbox{.}}{2013}]%
        {DitRGP13}
\bibfield{author}{\bibinfo{person}{Bogdan Dit}, \bibinfo{person}{Meghan
  Revelle}, \bibinfo{person}{Malcom Gethers}, {and} \bibinfo{person}{Denys
  Poshyvanyk}.} \bibinfo{year}{2013}\natexlab{}.
\newblock \showarticletitle{Feature location in source code: a taxonomy and
  survey}.
\newblock \bibinfo{journal}{\emph{Journal of Software: Evolution and Process}}
  \bibinfo{volume}{25}, \bibinfo{number}{1} (\bibinfo{year}{2013}),
  \bibinfo{pages}{53--95}.
\newblock


\bibitem[\protect\citeauthoryear{Ebert, Kuhrmann, and Prikladnicki}{Ebert
  et~al\mbox{.}}{2016a}]%
        {EbertKP16b}
\bibfield{author}{\bibinfo{person}{Christof Ebert}, \bibinfo{person}{Marco
  Kuhrmann}, {and} \bibinfo{person}{Rafael Prikladnicki}.}
  \bibinfo{year}{2016}\natexlab{a}.
\newblock \showarticletitle{Global Software Engineering: An Industry
  Perspective}.
\newblock \bibinfo{journal}{\emph{{IEEE} Software}} \bibinfo{volume}{33},
  \bibinfo{number}{1} (\bibinfo{year}{2016}), \bibinfo{pages}{105--108}.
\newblock


\bibitem[\protect\citeauthoryear{Ebert, Kuhrmann, and Prikladnicki}{Ebert
  et~al\mbox{.}}{2016b}]%
        {EbertKP16a}
\bibfield{author}{\bibinfo{person}{Christof Ebert}, \bibinfo{person}{Marco
  Kuhrmann}, {and} \bibinfo{person}{Rafael Prikladnicki}.}
  \bibinfo{year}{2016}\natexlab{b}.
\newblock \showarticletitle{Global Software Engineering: Evolution and Trends}.
  In \bibinfo{booktitle}{\emph{11th {IEEE} International Conference on Global
  Software Engineering, {ICGSE} 2016}}. \bibinfo{publisher}{{IEEE}},
  \bibinfo{pages}{144--153}.
\newblock


\bibitem[\protect\citeauthoryear{Girard and Girard}{Girard and Girard}{2015}]%
        {Girard15}
\bibfield{author}{\bibinfo{person}{John Girard} {and} \bibinfo{person}{JoAnn
  Girard}.} \bibinfo{year}{2015}\natexlab{}.
\newblock \showarticletitle{Defining knowledge management: Toward an applied
  compendium}.
\newblock \bibinfo{journal}{\emph{Online Journal of Applied Knowledge
  Management}} \bibinfo{volume}{3}, \bibinfo{number}{1} (\bibinfo{year}{2015}).
\newblock


\bibitem[\protect\citeauthoryear{Goel, Bhat, and Weber}{Goel
  et~al\mbox{.}}{2013}]%
        {GoelBW12}
\bibfield{author}{\bibinfo{person}{Sukriti Goel}, \bibinfo{person}{Jyoti~M.
  Bhat}, {and} \bibinfo{person}{Barbara Weber}.}
  \bibinfo{year}{2013}\natexlab{}.
\newblock \showarticletitle{End-to-End Process Extraction in Process Unaware
  Systems}. In \bibinfo{booktitle}{\emph{Business Process Management Workshops
  - {BPM} 2012 International Workshops. Revised Papers}}
  \emph{(\bibinfo{series}{Lecture Notes in Business Information Processing})},
  Vol.~\bibinfo{volume}{132}. \bibinfo{publisher}{Springer},
  \bibinfo{pages}{162--173}.
\newblock


\bibitem[\protect\citeauthoryear{Harman}{Harman}{2007}]%
        {mh:icpc-keynote}
\bibfield{author}{\bibinfo{person}{Mark Harman}.}
  \bibinfo{year}{2007}\natexlab{}.
\newblock \showarticletitle{Search Based Software Engineering for Program
  Comprehension}. In \bibinfo{booktitle}{\emph{15th International Conference on
  Program Comprehension {(ICPC} 2007)}}. \bibinfo{publisher}{{IEEE}},
  \bibinfo{pages}{3--13}.
\newblock


\bibitem[\protect\citeauthoryear{Harman, Gold, Hierons, and Binkley}{Harman
  et~al\mbox{.}}{2002}]%
        {mhetal:cosac}
\bibfield{author}{\bibinfo{person}{Mark Harman}, \bibinfo{person}{Nicolas
  Gold}, \bibinfo{person}{Robert~Mark Hierons}, {and} \bibinfo{person}{David
  Binkley}.} \bibinfo{year}{2002}\natexlab{}.
\newblock \showarticletitle{Code Extraction Algorithms which Unify Slicing and
  Concept Assignment}. In \bibinfo{booktitle}{\emph{{IEEE} Working Conference
  on Reverse Engineering ({WCRE 2002})}}. \bibinfo{publisher}{{IEEE}},
  \bibinfo{pages}{11 -- 21}.
\newblock


\bibitem[\protect\citeauthoryear{Hogan, Blomqvist, Cochez, d'Amato, de~Melo,
  Gutierrez, Gayo, Kirrane, Neumaier, Polleres, Navigli, Ngomo, Rashid, Rula,
  Schmelzeisen, Sequeda, Staab, and Zimmermann}{Hogan et~al\mbox{.}}{2020}]%
        {hogan2020knowledge}
\bibfield{author}{\bibinfo{person}{Aidan Hogan}, \bibinfo{person}{Eva
  Blomqvist}, \bibinfo{person}{Michael Cochez}, \bibinfo{person}{Claudia
  d'Amato}, \bibinfo{person}{Gerard de Melo}, \bibinfo{person}{Claudio
  Gutierrez}, \bibinfo{person}{José Emilio~Labra Gayo},
  \bibinfo{person}{Sabrina Kirrane}, \bibinfo{person}{Sebastian Neumaier},
  \bibinfo{person}{Axel Polleres}, \bibinfo{person}{Roberto Navigli},
  \bibinfo{person}{Axel-Cyrille~Ngonga Ngomo}, \bibinfo{person}{Sabbir~M.
  Rashid}, \bibinfo{person}{Anisa Rula}, \bibinfo{person}{Lukas Schmelzeisen},
  \bibinfo{person}{Juan Sequeda}, \bibinfo{person}{Steffen Staab}, {and}
  \bibinfo{person}{Antoine Zimmermann}.} \bibinfo{year}{2020}\natexlab{}.
\newblock \bibinfo{title}{Knowledge Graphs}.
\newblock
\newblock
\showeprint[arxiv]{2003.02320}~[cs.AI]


\bibitem[\protect\citeauthoryear{Jansen and Cusumano}{Jansen and
  Cusumano}{2012}]%
        {JansenC12}
\bibfield{author}{\bibinfo{person}{Slinger Jansen} {and}
  \bibinfo{person}{Michael~A. Cusumano}.} \bibinfo{year}{2012}\natexlab{}.
\newblock \showarticletitle{Defining Software Ecosystems: {A} Survey of
  Software Platforms and Business Network Governance}. In
  \bibinfo{booktitle}{\emph{Proceedings of the Forth International Workshop on
  Software Ecosystems}} \emph{(\bibinfo{series}{{CEUR} Workshop Proceedings})},
  Vol.~\bibinfo{volume}{879}. \bibinfo{publisher}{CEUR-WS.org},
  \bibinfo{pages}{40--58}.
\newblock
\urldef\tempurl%
\url{http://ceur-ws.org/Vol-879}
\showURL{%
\tempurl}


\bibitem[\protect\citeauthoryear{Jeong, Kim, Zimmermann, and Yi}{Jeong
  et~al\mbox{.}}{2009}]%
        {JeongLZY09}
\bibfield{author}{\bibinfo{person}{G. Jeong}, \bibinfo{person}{S. Kim},
  \bibinfo{person}{T. Zimmermann}, {and} \bibinfo{person}{K. Yi}.}
  \bibinfo{year}{2009}\natexlab{}.
\newblock \bibinfo{title}{Improving code review by predicting reviewers and
  acceptance of patches}.  (\bibinfo{year}{2009}),
  \bibinfo{numpages}{18}~pages.
\newblock
\newblock
\shownote{Research on Software Analysis for Error-free Computing Center
  Tech-Memo ROSAEC MEMO 2009-006.}


\bibitem[\protect\citeauthoryear{Kalgutkar, Kaur, Gonzalez, Stakhanova, and
  Matyukhina}{Kalgutkar et~al\mbox{.}}{2019}]%
        {KalgutkarKGSM19}
\bibfield{author}{\bibinfo{person}{Vaibhavi Kalgutkar},
  \bibinfo{person}{Ratinder Kaur}, \bibinfo{person}{Hugo Gonzalez},
  \bibinfo{person}{Natalia Stakhanova}, {and} \bibinfo{person}{Alina
  Matyukhina}.} \bibinfo{year}{2019}\natexlab{}.
\newblock \showarticletitle{Code Authorship Attribution: Methods and
  Challenges}.
\newblock \bibinfo{journal}{\emph{{ACM} Comput. Surv.}} \bibinfo{volume}{52},
  \bibinfo{number}{1} (\bibinfo{year}{2019}), \bibinfo{pages}{3:1--3:36}.
\newblock


\bibitem[\protect\citeauthoryear{Konat, Erdweg, and Visser}{Konat
  et~al\mbox{.}}{2018}]%
        {KonatEV18}
\bibfield{author}{\bibinfo{person}{Gabri{\"{e}}l Konat},
  \bibinfo{person}{Sebastian Erdweg}, {and} \bibinfo{person}{Eelco Visser}.}
  \bibinfo{year}{2018}\natexlab{}.
\newblock \showarticletitle{Scalable incremental building with dynamic task
  dependencies}. In \bibinfo{booktitle}{\emph{Proceedings of the 33rd
  {ACM/IEEE} International Conference on Automated Software Engineering, {ASE}
  2018}}. \bibinfo{publisher}{{ACM}}, \bibinfo{pages}{76--86}.
\newblock


\bibitem[\protect\citeauthoryear{Korel and Rilling}{Korel and Rilling}{1997}]%
        {korel:understanding}
\bibfield{author}{\bibinfo{person}{Bogdan Korel} {and} \bibinfo{person}{Jurgen
  Rilling}.} \bibinfo{year}{1997}\natexlab{}.
\newblock \showarticletitle{Dynamic Program Slicing in Understanding of Program
  Execution}. In \bibinfo{booktitle}{\emph{$5^{th}$ {IEEE} {I}nternational
  {W}orkshop on {P}rogram {C}omprenhesion ({IWPC'97})}}.
  \bibinfo{publisher}{{IEEE}}, \bibinfo{pages}{80--89}.
\newblock


\bibitem[\protect\citeauthoryear{L\"ammel, Kerber, and Praza}{L\"ammel
  et~al\mbox{.}}{2020}]%
        {workitem}
\bibfield{author}{\bibinfo{person}{Ralf L\"ammel}, \bibinfo{person}{Alvin
  Kerber}, {and} \bibinfo{person}{Liane Praza}.}
  \bibinfo{year}{2020}\natexlab{}.
\newblock \showarticletitle{Understanding What Software Engineers Are Working
  on -- The Work-Item Prediction Challenge}. In
  \bibinfo{booktitle}{\emph{Proceedings of the 28th {IEEE}/{ACM} International
  Conference on Program Comprehension (ICPC Industry Track)}}.
  \bibinfo{publisher}{{IEEE} / {ACM}}.
\newblock


\bibitem[\protect\citeauthoryear{Li, Sun, Leung, and Zhang}{Li
  et~al\mbox{.}}{2013}]%
        {LiSLZ13}
\bibfield{author}{\bibinfo{person}{Bixin Li}, \bibinfo{person}{Xiaobing Sun},
  \bibinfo{person}{Hareton Leung}, {and} \bibinfo{person}{Sai Zhang}.}
  \bibinfo{year}{2013}\natexlab{}.
\newblock \showarticletitle{A survey of code-based change impact analysis
  techniques}.
\newblock \bibinfo{journal}{\emph{Softw. Test., Verif. Reliab.}}
  \bibinfo{volume}{23}, \bibinfo{number}{8} (\bibinfo{year}{2013}),
  \bibinfo{pages}{613--646}.
\newblock


\bibitem[\protect\citeauthoryear{Lipcak and Rossi}{Lipcak and Rossi}{2018}]%
        {Lipcak018}
\bibfield{author}{\bibinfo{person}{Jakub Lipcak} {and} \bibinfo{person}{Bruno
  Rossi}.} \bibinfo{year}{2018}\natexlab{}.
\newblock \showarticletitle{A Large-Scale Study on Source Code Reviewer
  Recommendation}. In \bibinfo{booktitle}{\emph{44th Euromicro Conference on
  Software Engineering and Advanced Applications, {SEAA} 2018}}.
  \bibinfo{publisher}{{IEEE}}, \bibinfo{pages}{378--387}.
\newblock


\bibitem[\protect\citeauthoryear{Liu, He, Wu, and Li}{Liu
  et~al\mbox{.}}{2017}]%
        {LiuHWL17}
\bibfield{author}{\bibinfo{person}{Yaxin Liu}, \bibinfo{person}{Peng He},
  \bibinfo{person}{Gaoyan Wu}, {and} \bibinfo{person}{Yilu Li}.}
  \bibinfo{year}{2017}\natexlab{}.
\newblock \showarticletitle{Towards Understanding Developers' Collaborative
  Behavior in Open Source Software Ecosystems}.
\newblock \bibinfo{journal}{\emph{{JSW}}} \bibinfo{volume}{12},
  \bibinfo{number}{6} (\bibinfo{year}{2017}), \bibinfo{pages}{393--405}.
\newblock


\bibitem[\protect\citeauthoryear{Lundberg and Lee}{Lundberg and Lee}{2017}]%
        {LundbergL17}
\bibfield{author}{\bibinfo{person}{Scott~M. Lundberg} {and}
  \bibinfo{person}{Su{-}In Lee}.} \bibinfo{year}{2017}\natexlab{}.
\newblock \showarticletitle{A Unified Approach to Interpreting Model
  Predictions}. In \bibinfo{booktitle}{\emph{Advances in Neural Information
  Processing Systems 30: Annual Conference on Neural Information Processing
  Systems 2017}}. \bibinfo{publisher}{{NIPS}}, \bibinfo{pages}{4765--4774}.
\newblock


\bibitem[\protect\citeauthoryear{Lungu, Robbes, and Lanza}{Lungu
  et~al\mbox{.}}{2010}]%
        {LunguRL10}
\bibfield{author}{\bibinfo{person}{Mircea Lungu}, \bibinfo{person}{Romain
  Robbes}, {and} \bibinfo{person}{Michele Lanza}.}
  \bibinfo{year}{2010}\natexlab{}.
\newblock \showarticletitle{Recovering inter-project dependencies in software
  ecosystems}. In \bibinfo{booktitle}{\emph{{ASE} 2010, 25th {IEEE/ACM}
  International Conference on Automated Software Engineering}}.
  \bibinfo{publisher}{{ACM}}, \bibinfo{pages}{309--312}.
\newblock


\bibitem[\protect\citeauthoryear{Manikas}{Manikas}{2016}]%
        {Manikas16}
\bibfield{author}{\bibinfo{person}{Konstantinos Manikas}.}
  \bibinfo{year}{2016}\natexlab{}.
\newblock \showarticletitle{Revisiting software ecosystems Research: {A}
  longitudinal literature study}.
\newblock \bibinfo{journal}{\emph{Journal of Systems and Software}}
  \bibinfo{volume}{117} (\bibinfo{year}{2016}), \bibinfo{pages}{84--103}.
\newblock


\bibitem[\protect\citeauthoryear{Manikas and Hansen}{Manikas and
  Hansen}{2013}]%
        {ManikasH13}
\bibfield{author}{\bibinfo{person}{Konstantinos Manikas} {and}
  \bibinfo{person}{Klaus~Marius Hansen}.} \bibinfo{year}{2013}\natexlab{}.
\newblock \showarticletitle{Software ecosystems - {A} systematic literature
  review}.
\newblock \bibinfo{journal}{\emph{Journal of Systems and Software}}
  \bibinfo{volume}{86}, \bibinfo{number}{5} (\bibinfo{year}{2013}),
  \bibinfo{pages}{1294--1306}.
\newblock


\bibitem[\protect\citeauthoryear{Mao, Capra, Harman, and Jia}{Mao
  et~al\mbox{.}}{2017}]%
        {ke:jss-survey}
\bibfield{author}{\bibinfo{person}{Ke Mao}, \bibinfo{person}{Licia Capra},
  \bibinfo{person}{Mark Harman}, {and} \bibinfo{person}{Yue Jia}.}
  \bibinfo{year}{2017}\natexlab{}.
\newblock \showarticletitle{A survey of the use of crowdsourcing in software
  engineering}.
\newblock \bibinfo{journal}{\emph{Journal of Systems and Software}}
  \bibinfo{volume}{126} (\bibinfo{year}{2017}), \bibinfo{pages}{57--84}.
\newblock


\bibitem[\protect\citeauthoryear{{Motahari Nezhad}, Benatallah, Saint{-}Paul,
  Casati, and Andritsos}{{Motahari Nezhad} et~al\mbox{.}}{2008}]%
        {NezhadBSCA08}
\bibfield{author}{\bibinfo{person}{Hamid~R. {Motahari Nezhad}},
  \bibinfo{person}{Boualem Benatallah}, \bibinfo{person}{R{\'{e}}gis
  Saint{-}Paul}, \bibinfo{person}{Fabio Casati}, {and}
  \bibinfo{person}{Periklis Andritsos}.} \bibinfo{year}{2008}\natexlab{}.
\newblock \showarticletitle{Process spaceship: discovering and exploring
  process views from event logs in data spaces}.
\newblock \bibinfo{journal}{\emph{{PVLDB}}} \bibinfo{volume}{1},
  \bibinfo{number}{2} (\bibinfo{year}{2008}), \bibinfo{pages}{1412--1415}.
\newblock


\bibitem[\protect\citeauthoryear{Poshyvanyk, Marcus, Ferenc, and
  Gyim{\'{o}}thy}{Poshyvanyk et~al\mbox{.}}{2009}]%
        {PoshyvanykMFG09}
\bibfield{author}{\bibinfo{person}{Denys Poshyvanyk}, \bibinfo{person}{Andrian
  Marcus}, \bibinfo{person}{Rudolf Ferenc}, {and} \bibinfo{person}{Tibor
  Gyim{\'{o}}thy}.} \bibinfo{year}{2009}\natexlab{}.
\newblock \showarticletitle{Using information retrieval based coupling measures
  for impact analysis}.
\newblock \bibinfo{journal}{\emph{Empirical Software Engineering}}
  \bibinfo{volume}{14}, \bibinfo{number}{1} (\bibinfo{year}{2009}),
  \bibinfo{pages}{5--32}.
\newblock


\bibitem[\protect\citeauthoryear{Rahman, Roy, and Collins}{Rahman
  et~al\mbox{.}}{2016}]%
        {RahmanRC16}
\bibfield{author}{\bibinfo{person}{Mohammad~Masudur Rahman},
  \bibinfo{person}{Chanchal~K. Roy}, {and} \bibinfo{person}{Jason~A. Collins}.}
  \bibinfo{year}{2016}\natexlab{}.
\newblock \showarticletitle{CoRReCT: code reviewer recommendation in GitHub
  based on cross-project and technology experience}. In
  \bibinfo{booktitle}{\emph{Proceedings of the 38th International Conference on
  Software Engineering, {ICSE} 2016, Companion Volume}}.
  \bibinfo{publisher}{{ACM}}, \bibinfo{pages}{222--231}.
\newblock


\bibitem[\protect\citeauthoryear{Rempel, M{\"{a}}der, Kuschke, and
  Philippow}{Rempel et~al\mbox{.}}{2013}]%
        {RempelMKP13}
\bibfield{author}{\bibinfo{person}{Patrick Rempel}, \bibinfo{person}{Patrick
  M{\"{a}}der}, \bibinfo{person}{Tobias Kuschke}, {and} \bibinfo{person}{Ilka
  Philippow}.} \bibinfo{year}{2013}\natexlab{}.
\newblock \showarticletitle{Requirements Traceability across Organizational
  Boundaries - {A} Survey and Taxonomy}. In
  \bibinfo{booktitle}{\emph{Requirements Engineering: Foundation for Software
  Quality - 19th International Working Conference, {REFSQ} 2013}}
  \emph{(\bibinfo{series}{LNCS})}, Vol.~\bibinfo{volume}{7830}.
  \bibinfo{publisher}{Springer}, \bibinfo{pages}{125--140}.
\newblock


\bibitem[\protect\citeauthoryear{Ren, Ryder, St{\"o}rzer, and Tip}{Ren
  et~al\mbox{.}}{2005}]%
        {ren:chianti}
\bibfield{author}{\bibinfo{person}{Xiaoxia Ren}, \bibinfo{person}{Barbara~G.
  Ryder}, \bibinfo{person}{Maximilian St{\"o}rzer}, {and}
  \bibinfo{person}{Frank Tip}.} \bibinfo{year}{2005}\natexlab{}.
\newblock \showarticletitle{Chianti: a change impact analysis tool for {J}ava
  programs}. In \bibinfo{booktitle}{\emph{27th International Conference on
  Software Engineering ({ICSE} 2005)}}. \bibinfo{publisher}{ACM},
  \bibinfo{pages}{664--665}.
\newblock


\bibitem[\protect\citeauthoryear{Ribeiro, Singh, and Guestrin}{Ribeiro
  et~al\mbox{.}}{2016}]%
        {Ribeiro0G16}
\bibfield{author}{\bibinfo{person}{Marco~T{\'{u}}lio Ribeiro},
  \bibinfo{person}{Sameer Singh}, {and} \bibinfo{person}{Carlos Guestrin}.}
  \bibinfo{year}{2016}\natexlab{}.
\newblock \showarticletitle{"Why Should {I} Trust You?": Explaining the
  Predictions of Any Classifier}. In \bibinfo{booktitle}{\emph{Proceedings of
  the 22nd {ACM} {SIGKDD} International Conference on Knowledge Discovery and
  Data Mining}}. \bibinfo{publisher}{{ACM}}, \bibinfo{pages}{1135--1144}.
\newblock


\bibitem[\protect\citeauthoryear{Rudin}{Rudin}{2019}]%
        {Rudin19}
\bibfield{author}{\bibinfo{person}{Cynthia Rudin}.}
  \bibinfo{year}{2019}\natexlab{}.
\newblock \showarticletitle{Stop explaining black box machine learning models
  for high stakes decisions and use interpretable models instead}.
\newblock \bibinfo{journal}{\emph{Nature Machine Intelligence}}
  \bibinfo{volume}{1} (\bibinfo{date}{05} \bibinfo{year}{2019}),
  \bibinfo{pages}{206--215}.
\newblock


\bibitem[\protect\citeauthoryear{Rudin and Ustun}{Rudin and Ustun}{2018}]%
        {RudinU18}
\bibfield{author}{\bibinfo{person}{Cynthia Rudin} {and} \bibinfo{person}{Berk
  Ustun}.} \bibinfo{year}{2018}\natexlab{}.
\newblock \showarticletitle{Optimized Scoring Systems: Toward Trust in Machine
  Learning for Healthcare and Criminal Justice}.
\newblock \bibinfo{journal}{\emph{Interfaces}} \bibinfo{volume}{48},
  \bibinfo{number}{5} (\bibinfo{year}{2018}), \bibinfo{pages}{449--466}.
\newblock


\bibitem[\protect\citeauthoryear{Silva}{Silva}{2012}]%
        {silva:slice-vocab}
\bibfield{author}{\bibinfo{person}{Josep Silva}.}
  \bibinfo{year}{2012}\natexlab{}.
\newblock \showarticletitle{A Vocabulary of Program Slicing-Based Techniques}.
\newblock \bibinfo{journal}{\emph{Comput. Surveys}} \bibinfo{volume}{44},
  \bibinfo{number}{3} (\bibinfo{date}{June} \bibinfo{year}{2012}),
  \bibinfo{pages}{12:1 -- 12:48}.
\newblock


\bibitem[\protect\citeauthoryear{Sun, Gao, Du, and Ye}{Sun
  et~al\mbox{.}}{2016}]%
        {SunGDY16}
\bibfield{author}{\bibinfo{person}{Xuan Sun}, \bibinfo{person}{Xin Gao},
  \bibinfo{person}{Huiying Du}, {and} \bibinfo{person}{Wei Ye}.}
  \bibinfo{year}{2016}\natexlab{}.
\newblock \showarticletitle{A Query Language of Data Provenance Based on
  Dependency View for Process Analysis}. In \bibinfo{booktitle}{\emph{The 28th
  International Conference on Software Engineering and Knowledge Engineering,
  {SEKE} 2016}}. \bibinfo{publisher}{{KSI} Research Inc. and Knowledge Systems
  Institute Graduate School}, \bibinfo{pages}{110--113}.
\newblock


\bibitem[\protect\citeauthoryear{Wang and Redmiles}{Wang and Redmiles}{2016}]%
        {WangR16}
\bibfield{author}{\bibinfo{person}{Yi Wang} {and} \bibinfo{person}{David~F.
  Redmiles}.} \bibinfo{year}{2016}\natexlab{}.
\newblock \showarticletitle{Cheap talk, cooperation, and trust in global
  software engineering - An evolutionary game theory model with empirical
  support}.
\newblock \bibinfo{journal}{\emph{Empirical Software Engineering}}
  \bibinfo{volume}{21}, \bibinfo{number}{6} (\bibinfo{year}{2016}),
  \bibinfo{pages}{2233--2267}.
\newblock


\bibitem[\protect\citeauthoryear{Weiser}{Weiser}{1979}]%
        {weiser:slicing79}
\bibfield{author}{\bibinfo{person}{Mark Weiser}.}
  \bibinfo{year}{1979}\natexlab{}.
\newblock \emph{\bibinfo{title}{Program slices: Formal, psychological, and
  practical investigations of an automatic program abstraction method}}.
\newblock \bibinfo{thesistype}{Ph.D. Dissertation}. \bibinfo{school}{University
  of Michigan, Ann Arbor, MI}.
\newblock


\bibitem[\protect\citeauthoryear{Yu, Wang, Yin, and Wang}{Yu
  et~al\mbox{.}}{2016}]%
        {YuWYW16}
\bibfield{author}{\bibinfo{person}{Yue Yu}, \bibinfo{person}{Huaimin Wang},
  \bibinfo{person}{Gang Yin}, {and} \bibinfo{person}{Tao Wang}.}
  \bibinfo{year}{2016}\natexlab{}.
\newblock \showarticletitle{Reviewer recommendation for pull-requests in
  GitHub: What can we learn from code review and bug assignment?}
\newblock \bibinfo{journal}{\emph{Information {\&} Software Technology}}
  \bibinfo{volume}{74} (\bibinfo{year}{2016}), \bibinfo{pages}{204--218}.
\newblock


\end{thebibliography}
\bibliographystyle{ACM-Reference-Format}

\end{document}